\documentclass{pasj01}
\Received{$\langle$reception date$\rangle$}
\Accepted{$\langle$acceptance date$\rangle$}
\Published{$\langle$publication date$\rangle$}
\usepackage{multirow}
\usepackage{subfigure}
\usepackage{natbib}
\usepackage{url}
\usepackage[usenames,dvipsnames]{xcolor}
\usepackage{ulem}
\usepackage[T1]{fontenc}
\usepackage[english]{babel}

\begin{document}

\title{Radio-Infrared Correlation for Local Dusty Galaxies and Dusty AGNs from the AKARI All Sky Survey}
\author{A. Solarz$^1$, A. Pollo$^{1,2}$, M. Bilicki$^{3,4}$, A. P{\c e}piak$^{2}$, T. T. Takeuchi$^{5}$, P. Pi{\c a}tek$^2$}%
\altaffiltext{}{\\$^1$National Center for Nuclear Research, ul.\ Ho\.{z}a 69, 00-681 Warsaw, Poland \\$^2$Astronomical Observatory of the Jagiellonian University, ul. Orla 171, 30-244 Krak\'{o}w, Poland \\ $^3$Leiden Observatory, Leiden University, P.O.Box 9513, 2300RA Leiden, The Netherlands \\
$^4$Center for Theoretical Physics, Polish Academy of Sciences, al. Lotnik\'{o}w 32/46, 02-668, Warsaw, Poland \\
$^5$Division of Astroparticle and Astrophysical Science, Graduate School of Science, Nagoya University, Furo-Cho, Chikusa-ku, Nagoya 464-8602 Japan}
\email{aleksandra.solarz@ncbj.gov.pl}


\maketitle

\begin{abstract}
We use the new release of the AKARI Far-Infrared all sky Survey matched with the NVSS radio database to investigate the local ($z<0.25$) far infrared-radio correlation (FIRC) of different types of extragalactic sources. To obtain the redshift information for the AKARI FIS sources we crossmatch the catalogue with the SDSS DR8. This also allows us to use emission line properties to divide sources into four categories: i) star-forming galaxies (SFGs), ii) composite galaxies (displaying both star-formation and active nucleus components), iii) Seyfert galaxies, and iv) low-ionization nuclear emission-line region (LINER) galaxies. 
 We find that the Seyfert galaxies have the lowest FIR/radio flux ratios and display excess radio emission when compared to the SFGs. 
We conclude that FIRC can be used to separate SFGs and AGNs only for the most radio-loud objects.
\end{abstract}

\section{Introduction}


A tight far infrared-radio correlation (FIRC) for galaxies is a well known phenomenon, which appears to extend from the local Universe (\citealt{helou85,condon92,yun01, bell03, mauch07}) up to high redshifts (\citealt{garrett02,apple04,chapman05,kovacs06,sajina08,ibar08,sargent10,michalowski10,ivison10,magnelli15,delhaize17,molnar18}) and for a wide range of galaxy masses and types \citep{condon91}. It is  
 attributed to recent star formation processes which produce both far infrared (FIR) and synchrotron radiation.
 FIR radiation is related to the star-formation  process in galaxies through dust particles produced througout the life-cycle of the galaxy, which  absorb the ultraviolet radiation emitted by young, massive, hot stars, and re-emit it at longer wavelengths.
Location of the FIR radiation peak varies significantly depending on the intensity of the interstellar radiation field (from $\sim$ 150 $\mu$m
in the coolest galaxies to $\sim$ 60 $\mu$m or shorter in the warmest ones), and it is a good indicator of dust heating in the interstellar medium (e.g. \citealt{cunha08}).
Once the massive stars explode as supernovae, cosmic rays are accelerated in the magnetic field of the galaxy producing synchrotron emission. This
 emission is characterized by a power-law spectrum, which makes it the dominant component of galaxy spectra in the $1 <\nu< 10$ GHz ($30 >\lambda> 3$ cm) range (\citealt{condon02,bressan02,bell03,clemens10,murphy11,mancuso17}).

FIRC is observed both in star-forming galaxies (SFGs) and in a large fraction of sources with an active galactic nucleus (AGN), and a similar trend is also seen for the mid-IR (MIR)-radio correlation (\citealt{donley05}, \citealt{park08}). Past observations indicated that AGNs display a larger scatter in the FIRC than SFGs (e.g. \citealt{condon82}),  
which was therefore considered useful in selecting obscured type 2 AGNs with a radio-loud nature, missed by deep X-ray observations (\citealt{yun01,condon02,ibar08}). 
However, more recent studies show that optically selected AGNs often follow the correlation as tightly as SFGs (\citealt{obric06,mauch07}). In particular, \cite{moric10},  based on a large cross-correlated sample of  the NRAO VLA Sky Survey (NVSS), Infrared Astronomical Satellite (IRAS) and Sloan Digital Sky Survey (SDSS) data, found that most galaxies containing AGNs have FIR/radio flux ratios indistinguishable from those of the SFGs that define the FIRC, with the exception of Seyfert galaxies. The latter show lower FIR/radio flux ratios, albeit still within the scatter of SFGs.

In this work we present a new study of the FIRC based on the newest measurements from the AKARI Far-Infrared All Sky Survey cross-correlated with NVSS.
The AKARI data provide better sensitivities than the preceding all-sky infrared mission of the IRAS satellite. While IRAS reached 1~Jy at 100~$\mu$m, AKARI detections go down to 0.2 Jy in the 90~$\mu$m passband (\citealt{yamamura10}). In addition, the AKARI All Sky Survey has much larger sky coverage (exceeding 98\% of the sky; \citealt{nakagawa07}) than targeted observations provided by \textit{Spitzer} \citep{spitzer} and \textit{Herschel} Space Observatories, which observed $\sim 10\%$ of the whole sky\footnote{Based on data products available in the NASA/IPAC Infrared Science Archive; \url{https://irsa.ipac.caltech.edu/Missions/herschel.html}.}.

\section{Data}

The Far Infrared Surveyor (FIS) instrument onboard the AKARI satellite scanned 94\% of the entire sky at least twice in the 16 months of the cryogenic mission phase, which led to the all-sky catalog covering four photometric far-infrared bands centered at 60 $\mu $m, 90 $\mu $m, 140 $\mu $m and 160 $\mu $m (the \textit{N60}, \textit{WIDE-S}, \textit{WIDE-L} and \textit{N160} bands, respectively; \citealt{kawada07}).

The newest version of the AKARI All Sky Survey in the FIR contains $950,365$ sources, out of which $410,623$ (43\%) 
were detected in all four passbands.
The point spread function of the AKARI FIR bands is $37\pm1"$, $39\pm1"$, $58\pm3"$ and $61\pm4"$ at 60, 90, 140 and 160~$\mu$m, respectively (\citealt{kawada07}, \citealt{doi15}, \citealt{takita15}).
To ensure good quality of photometric measurements for our analysis we chose those objects which have the flux quality indicator FQUAL=3 at 90~$\mu$m \citep{yamamura10}.
As mentioned above, the depth of the catalogue at 90~$\mathrm{\mu}$m reaches $\sim 0.2$ Jy at the 3$\sigma$ level.

Following \citet{pollo10} we selected objects lying in regions of low Galactic emission at 100~$\mathrm{\mu}$m, based on the \cite{schlegel98} maps. 
We restricted our analysis to sources with line-of-sight Galactic emission lower than 5~MJy~$\mathrm{sr^{-1}}$, which gave us a catalogue of $61,576$ galaxies (6\% of the original dataset). 

We performed a crossmatch between the AKARI-FIS catalogue and the NVSS radio sample \citep{condon98} within a $15"$ matching radius (e.g. \citealt{pepiak14}). This step left us with $20,274$ objects available for the FIRC analysis. To obtain redshift information for these sources we further crossmatched the data with the SDSS DR8\footnote{\url{https://www.sdss.org/dr12/spectro/galaxy_mpajhu/}}. 
We restricted ourselves to the SDSS DR8 as it offers line ratio diagnostics which allow for the exact determination of the nature of the 
sources (\citealt{kauff03,kewley01,kewley06}). 
The crossmatch AKARI-FIS $\times$ SDSS DR8 $\times$ NVSS resulted in a sample of $6,181$ objects, all of them with a redshift measurement.
\begin{figure}
  \centering
  \includegraphics[width=0.7\linewidth]{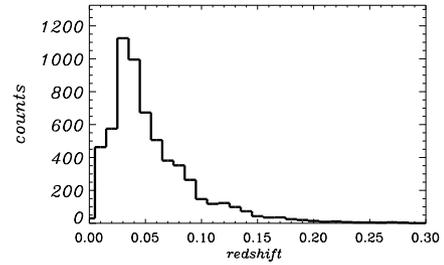}
  \caption{Redshift distribution of all the sources in the final catalogue of AKARI-FIS x SDSS DR8 x NVSS.}
\label{zdist}
\end{figure}%
Figure~\ref{zdist} shows the redshift distribution of the final catalogue. The distribution has a primary peak located at $z\sim 0.03$ with a possible secondary peak emerging at $z\sim0.12$. 

\begin{figure*}
  \centering
  \includegraphics[width=0.7\linewidth]{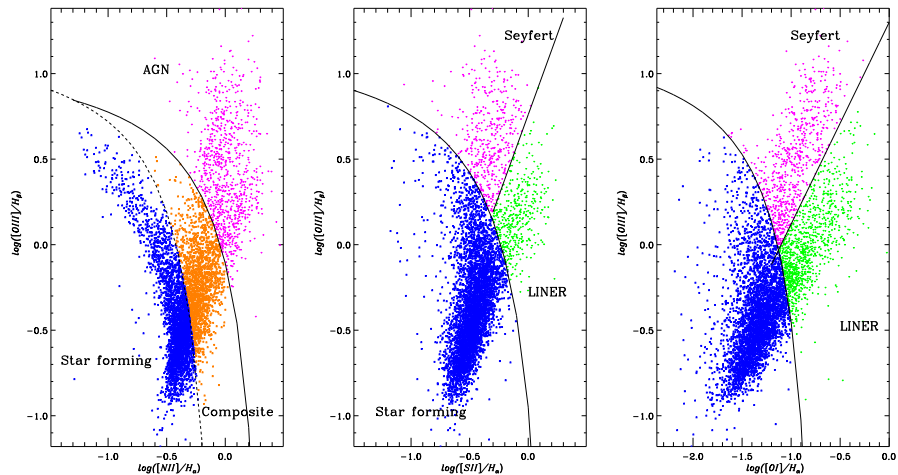}
  \caption{BPT spectroscopic diagnostic diagrams of our AKARI x NVSS x SDSS DR8 sample used to separate star-forming galaxies (marked by blue points), composite galaxies (marked by orange points), Seyfert galaxies (marked by pink points) and LINERs (marked by green points).}
\label{bpt}
\end{figure*}

A common diagnostic tool used to distinguish between AGNs and SFGs is the so-called BPT diagram \citep{BPT}. We present the BPT diagram of our sample in Fig.~\ref{bpt}. Following the selection criteria of \cite{kauff03} and \cite{kewley01, kewley06}  we find that $2,813$ (45\%) of our sample are SFGs, $2,666$ (43\%) are composite galaxies (containing contributions from both AGN and
star formation) and 629 (10\%) are AGNs, including 392 Seyfert galaxies and 237 low-ionization nuclear emission-line regions (LINER galaxies). 
In addition, 40 of the galaxies are classified as Seyferts based on either $[SII]/H\alpha$  or $[OI]/H\alpha$ diagrams but classified as LINERs based on $[OI]/H\alpha$ or $[SII]/H\alpha$ diagrams. Moreover, we find 33 galaxies classified as composite based on the $[NII]/H\alpha$ diagram, but shifted into the AGN region on either the $[SII]/H\alpha$  or $[OI]/H\alpha$ diagram.  
 These ambiguous sources are not included in the subsequent analysis.

\section{Far Infrared-Radio Correlation of AKARI galaxies}

As we have sources with redshifts measured, we compute their absolute far-infrared luminosities $L_{FIR}$ as 

\begin{equation}
L_{FIR}=\Delta \nu(\mathrm{WIDE-S})L_{\nu}(90~\mathrm{\mu} m)+ 
\end{equation}
\begin{center}
$+\Delta \nu(\mathrm{WIDE-L})L_{\nu}(140~\mathrm{\mu} m).$
\end{center}

with $\Delta \nu(\mbox{WIDE-S})=1.47 \times 10^{12}$[Hz],
$\Delta \nu(\mbox{WIDE-L})=0.831\times 10^{12}$[Hz], and
${L_{\nu}=4\pi D^{2}_{L_{\nu}} S_{\nu}}$, where $S_{\nu}$ is the flux density measured for each source through a specific AKARI filter, while $D_{L}$ is the luminosity distance. 
The equations were derived by comparing the AKARI FIS data with the IRAS all-sky survey. As \citet{takeuchi10} reported, the \textit{N60} filter sensitivity is
lower than that of the wide bands, and therefore it is beneficial to estimate IR luminosity by using only the WIDE-S and WIDE-L bands. 
We calculate the radio luminosities as:
\begin{equation}
\mathrm{\mathit{L_{1.4GHz}}=\frac{4\pi \mathit{D^{2}_{L}}}{(1+z)^{1-\alpha}} \mathit{S_{1.4GHz}}},
\end{equation}
where $S_{1.4GHz}$ is the radio flux as measured by NVSS and $\alpha$ is the radio spectral index. We assume $\alpha=0.7$ following \citet{condon92}, who reported that SFGs usually have a mean spectral index varying between $0.8$ and $0.7$ with a dispersion of 0.24, and \citet{lian83} who measured $\alpha$ for galaxies with nuclear activity to be $\sim 0.75$.

\begin{figure*}
  \centering
  \includegraphics[width=.9\linewidth]{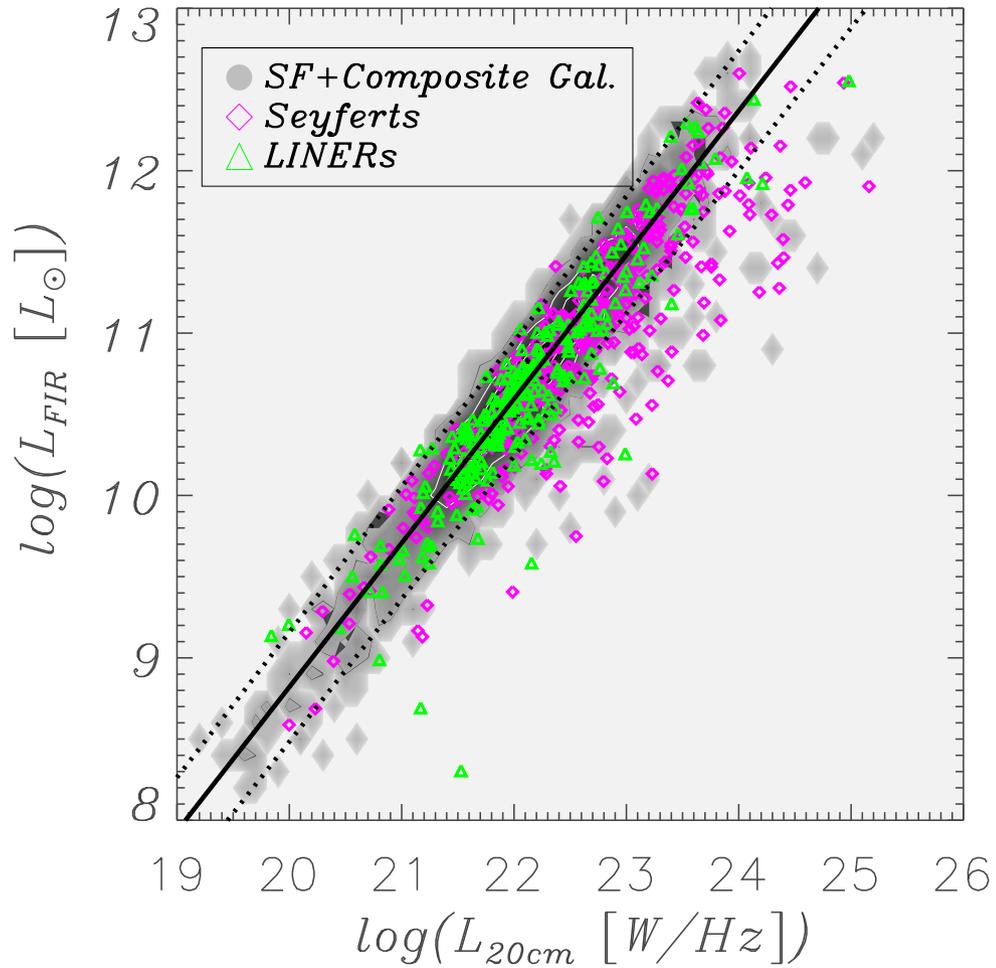}
  \caption{Total infrared luminosity ($L_{FIR}$) plotted against radio luminosity ($L_{1.4 GHz}$). The solid line represents the best least-square fit to the data, and the dashed lines mark the 95\% confidence interval of the linear regression. Green triangles represent LINERs, pink diamonds represent Seyfert galaxies, and shaded contours represent SF and composite galaxies.}
\label{ltirlradio}
\end{figure*}

The FIRC for the four main classes of objects: SF, composite, Seyfert and LINER galaxies  is shown in Fig.~\ref{ltirlradio} together with the best least-square fit to the data. To obtain the relationship between $L_{FIR}$ and $L_{1.4GHz}$ we fitted the following model:
$y=a∗x+b$, where $y=\log(L_{FIR})$ and $x=\log(L_{1.4 GHz})$. 
 We found tight correlation for all the categories of objects, with the best-fit parameters $a=0.887\pm0.004$ and $b=-8.920\pm 0.091$.

\begin{figure*}
  \centering
  \begin{subfigure}{\includegraphics[scale=0.4]{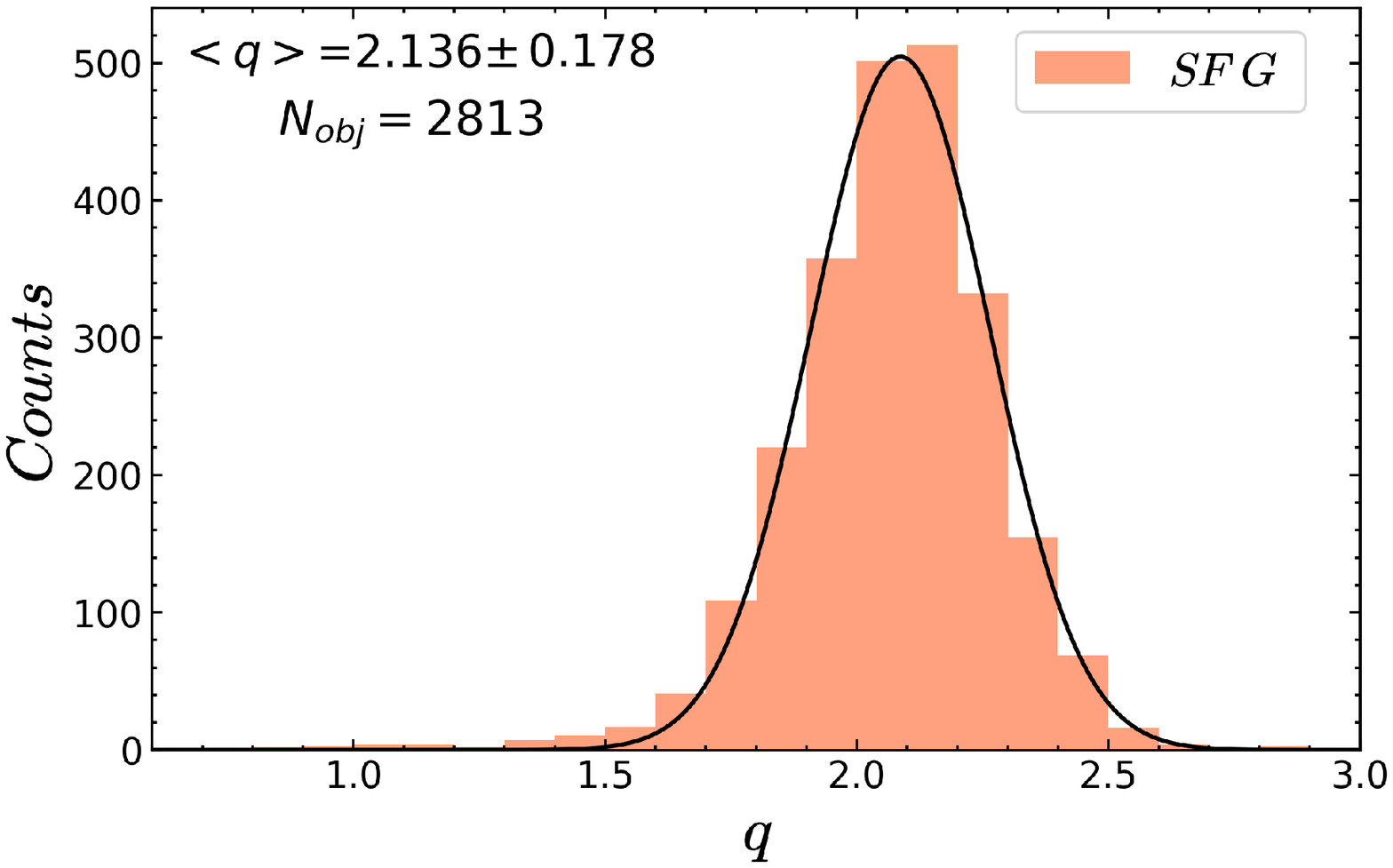}}
\end{subfigure}
  \begin{subfigure}{\includegraphics[scale=0.4]{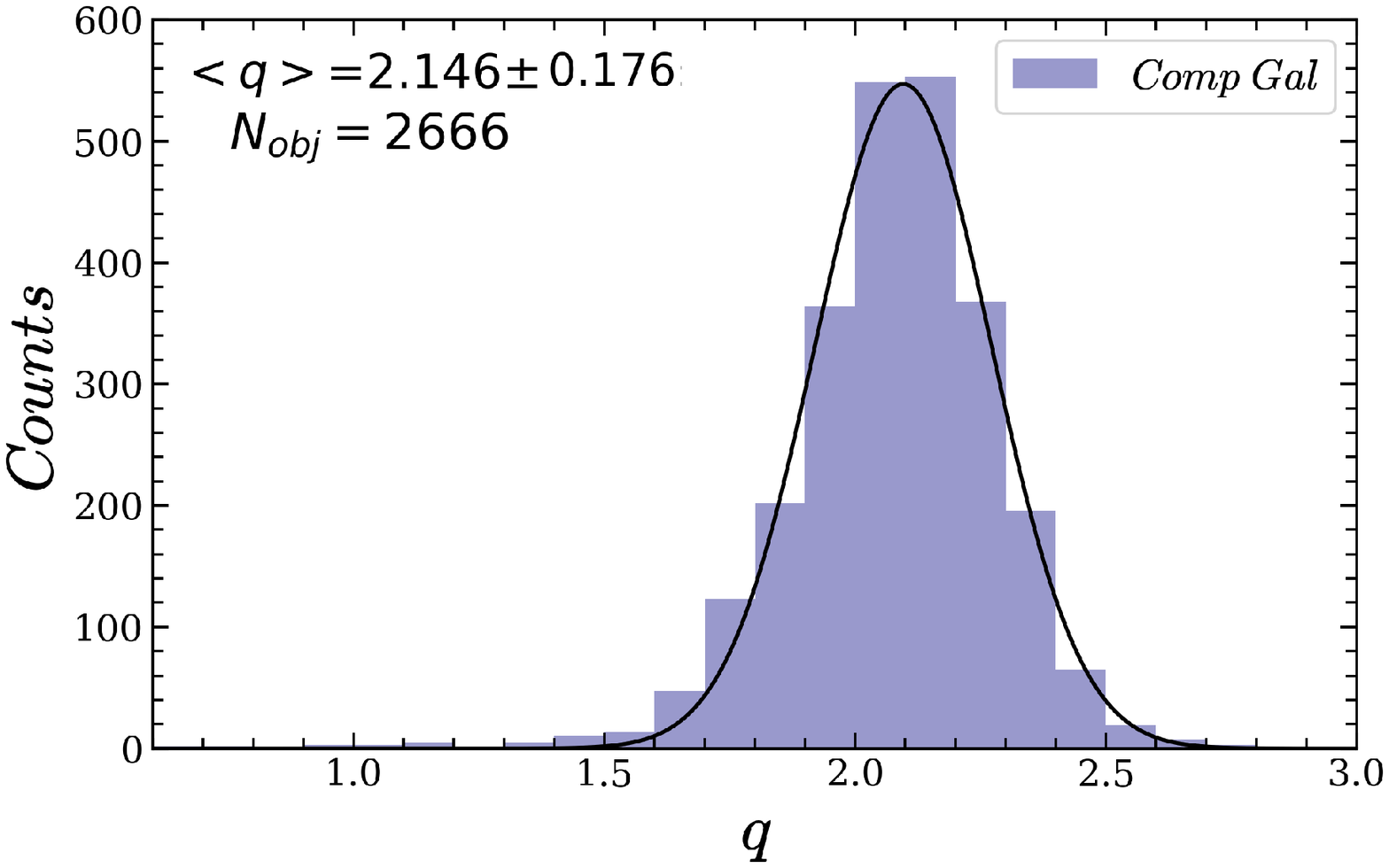}}
\end{subfigure}
  \begin{subfigure}{\includegraphics[scale=0.4]{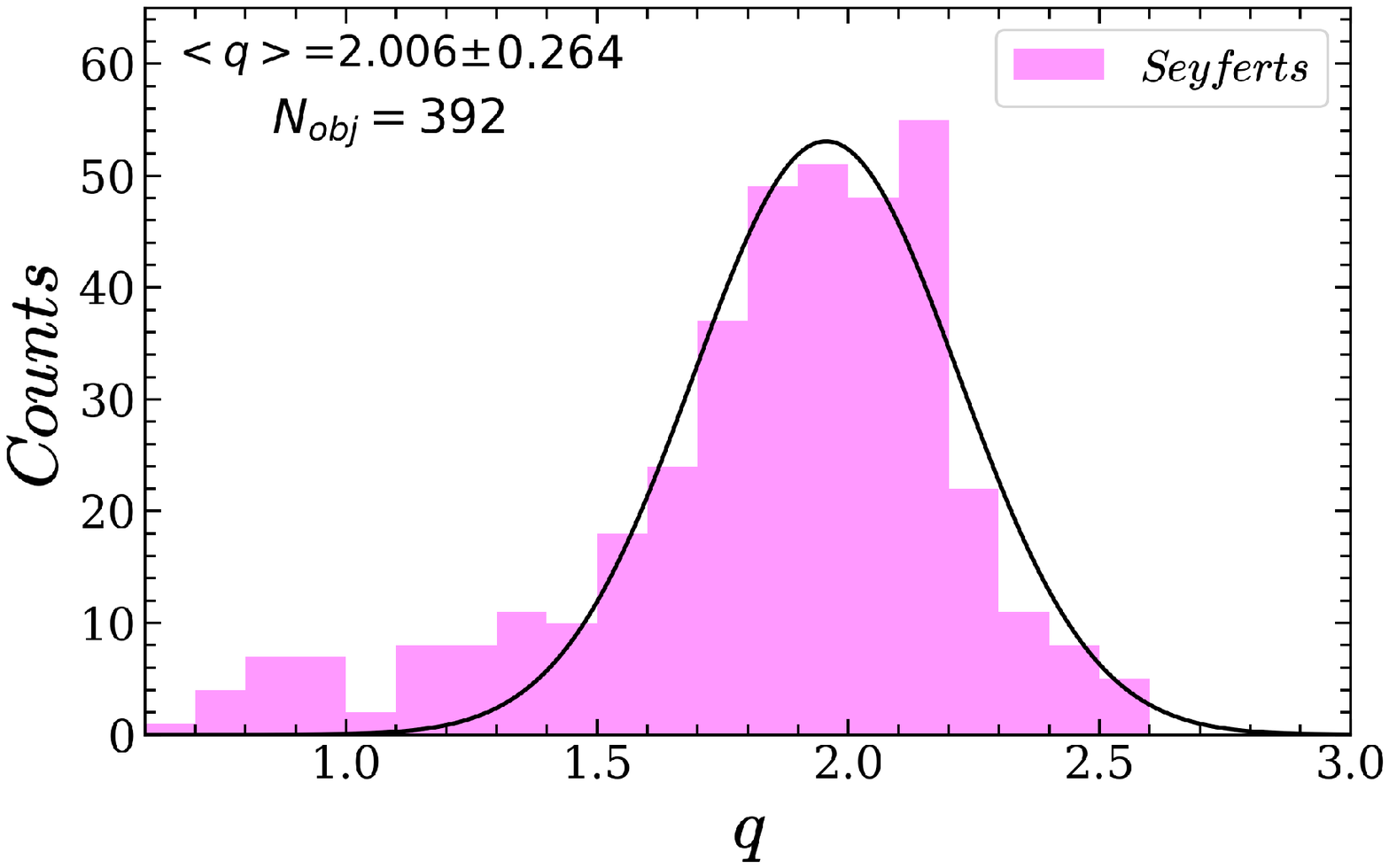}}
\end{subfigure} 
\begin{subfigure}{\includegraphics[scale=0.4]{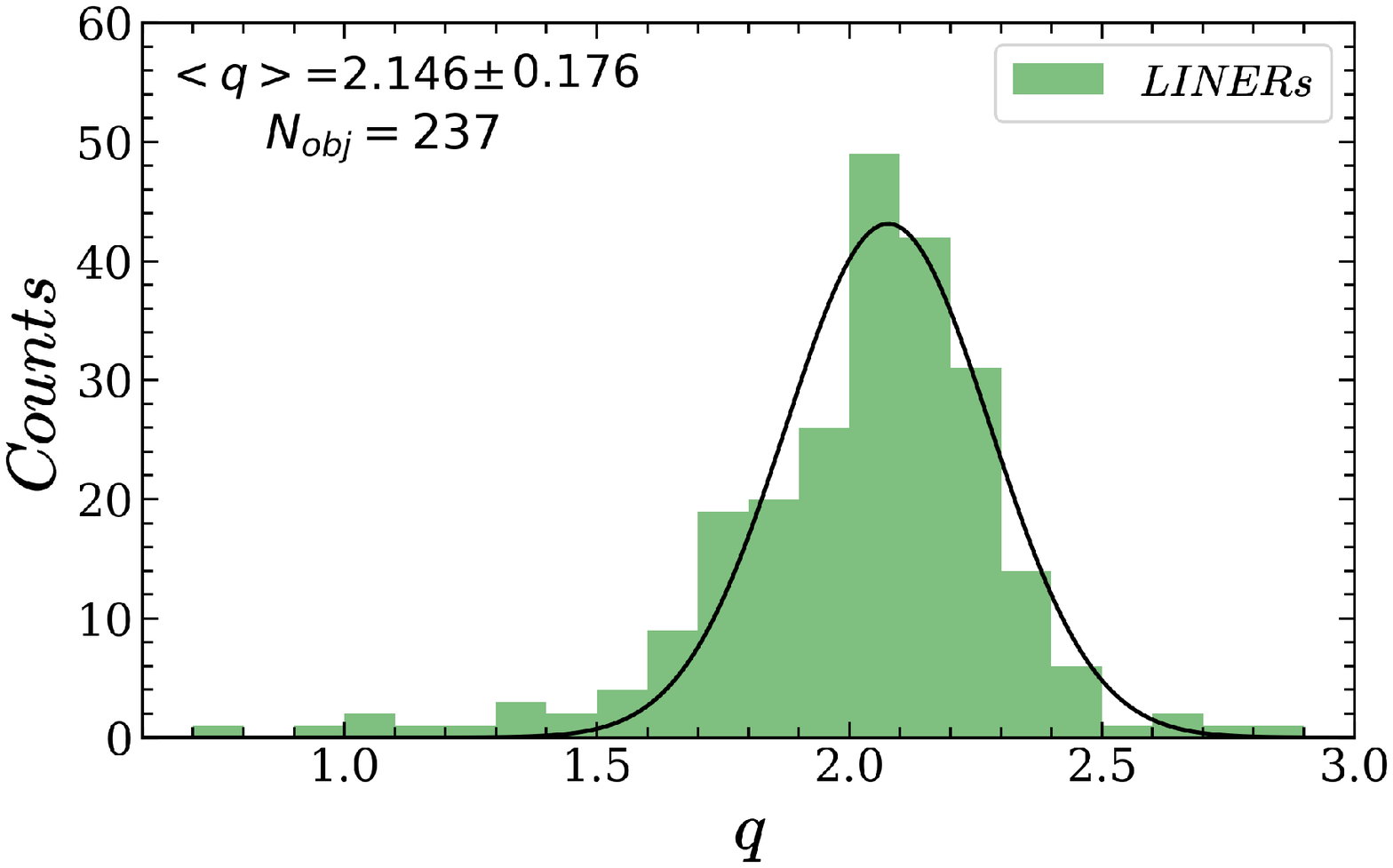}}
\end{subfigure}
  \caption{Distribution of the $q$ parameter for each considered source population: SFG (top left panel), composite galaxies (top right panel), Seyfert galaxies (bottom left panel) and LINERs (bottom right panel). Solid lines mark Gaussian fits. The number of objects in the sample ($N_{obj}$), the $<q>$ and the dispersion ($\sigma$) are indicated in the plots.}
\label{nq}
\end{figure*}

Following \cite{helou85} we calculate the `$q$' parameter to quantify the FIRC:
\begin{equation}
q=\log\left(\frac{L_{FIR} [\mathrm{W}]}{3.75\times 10^{12}}\right)-\log(L_{1.4GHz} [\mathrm{W\,Hz^{-1}}]).
\end{equation}
The mean $q$ ($\langle q \rangle$) parameter for the whole AKARI FIS $\times$ NVSS $\times$ SDSS DR8 dataset is $\langle q \rangle =2.04\pm 0.27$. 
We measure $\langle q \rangle=2.01\pm0.26$ for Seyfert galaxies, while for SFGs, composites and LINERs we find $\langle q \rangle\approx 2.14\pm0.18$ (see Table~\ref{sum}). 
Figure \ref{nq} shows the distribution of $q$ for all the considered source populations. 
Distributions of $q$ for SF and composite galaxies are very similar, and they have the same means; the latter is true also for LINERs.
The distribution for Seyferts, on the other hand, is characterized by a longer tail towards lower values of $q$.
SFGs, composites, and to a lesser extent also LINERs, seem to be Gaussian-distributed, while for Seyfert galaxies we observe a clearly skewed distribution. The latter is expected: Seyferts
show a larger scatter in $q$ when compared with the other considered source populations, with many sources on the high radio-luminosity side. 
However,  similar behaviour is also observed for some sources in the other classes.  
We set the separation between `radio-normal' and `radio-excess' sources at $q=1.5$, corresponding to a 2$\sigma$ deviation from the peak of the $q$ distribution for all the objects in our sample. We find 64 Seyfert galaxies (16\% of the full Seyfert sample) with radio excess. In contrast, 12 LINERs (5\% of the total LINER sample), 38 composite galaxies (1.4\% of the full composite galaxy sample) and 37 SFGs (1.3\% of the full SFG sample) show radio excess.

\begin{figure}
  \centering
  \includegraphics[width=.9\linewidth]{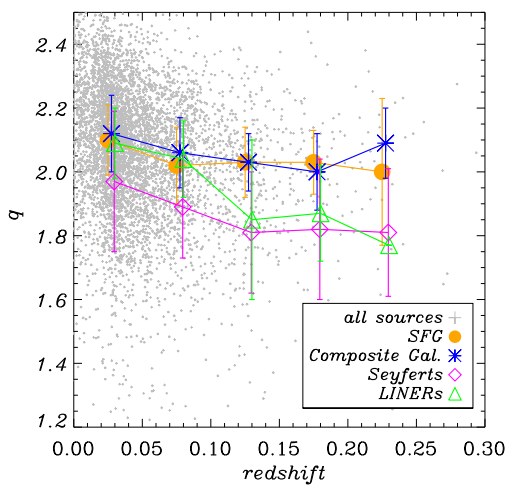}
  \caption{Distribution of the $q$ parameter for each considered source population: SFG (orange circles), composite galaxies (blue asterisks), Seyfert galaxies (pink diamonds) and LINERs (green triangles), as a function of redshift $z$. We calculate the mean $q$  and its standard deviation in redshift bins  with a $\Delta z=0.05$ step. Grey points mark the position of the whole sample of AKARI FIS sources.}
\label{qz}
\end{figure}

\begin{figure}
  \centering
  \includegraphics[width=.9\linewidth]{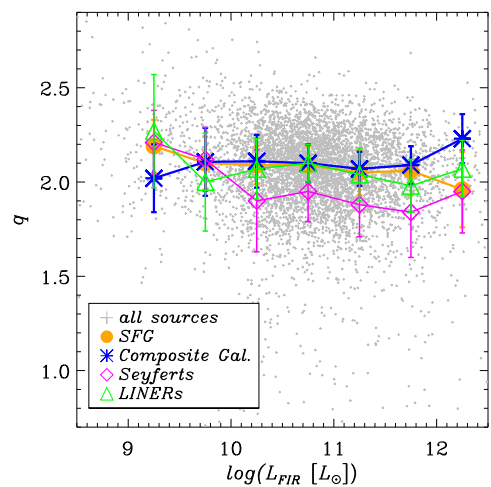}
  \caption{Distribution of the $q$ parameter for each considered source population as a function of $L_{FIR}$. Mean $q$ is calculated in $L_{FIR}$ logarithmic bins of 0.5. Points for different source populations are the same as in Fig.~\ref{qz}.}
\label{qltir}
\end{figure}

\begin{figure}
  \centering
  \includegraphics[width=.9\linewidth]{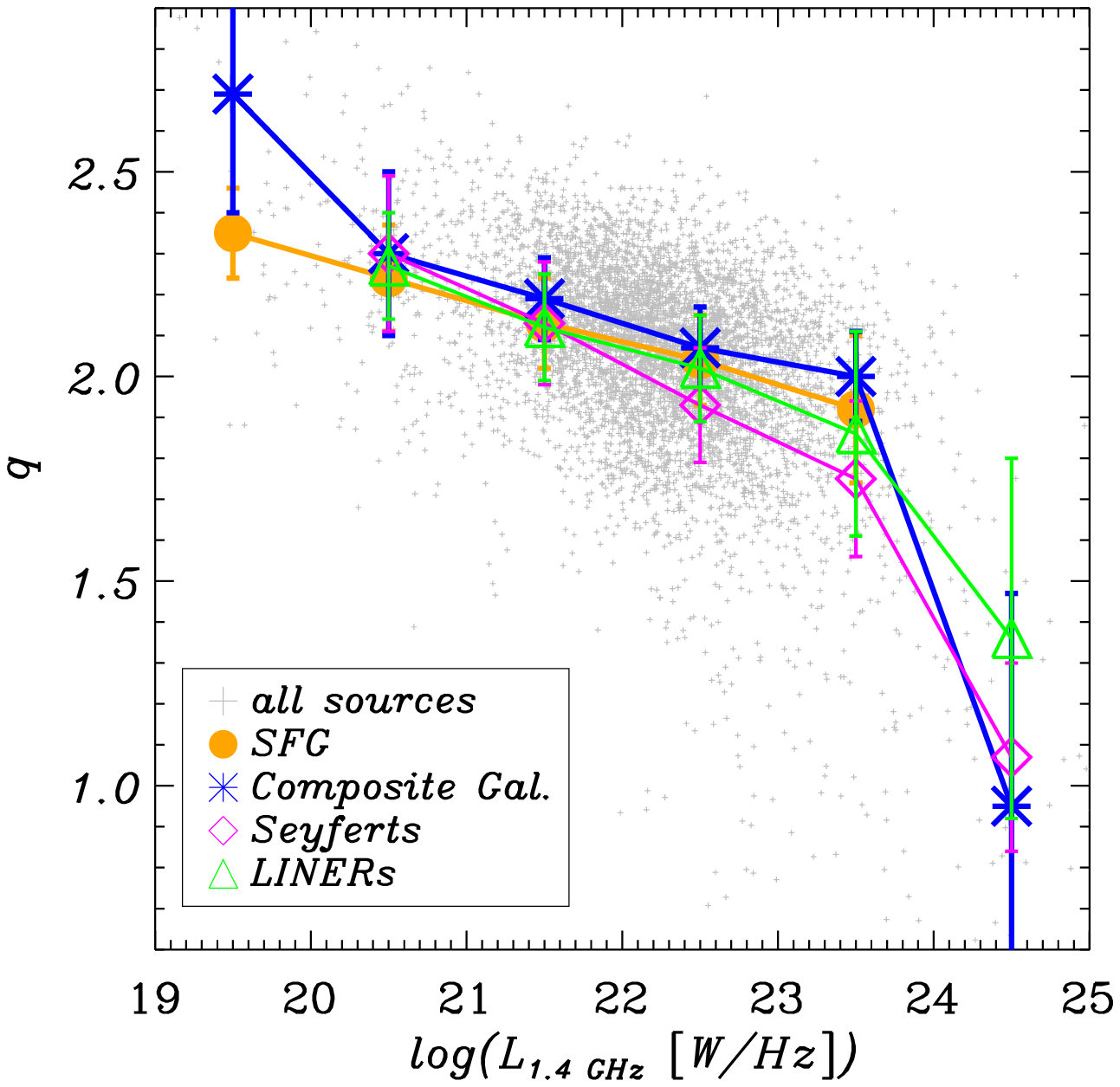}
  \caption{Distribution of the $q$ parameter for each considered source population as a function of $L_{1.4 GHz}$. Mean $q$ is calculated in $L_{1.4 GHz}$ logarithmic bins of 1. Points for different source populations are the same as in Fig.~\ref{qz}. }
\label{qlradio}
\end{figure}

In order to verify the differences between different populations of galaxies we have applied the Kolmogorov-Smirnov (K--S) test.
The distribution of $q$ for the total sample of composite galaxies is indistinguishable from that for SFGs, with p-value=0.252 of the null-hypothesis being true. However, we find that there is a statistical difference between the total sample of SFGs and the two AGN types (with p-value $< 2.2 \cdot 10^{-16}$ for SFGs vs. Seyfert galaxies and with p-value=0.010 for SFGs vs. LINERS).
We additionally separate our galaxy samples into several redshift and luminosity bins (both $L_{FIR}$ and $L_{1.4GHz}$)\footnote{In what follows, $L_{FIR}$ will be given in units of $L_{\odot}$, while $L_{1.4 GHz}$ in W/Hz.}, where we find different results (see Tables~\ref{ksz},~\ref{ksltir},~\ref{kslradio} and Figues~\ref{qz}, \ref{qltir}, \ref{qlradio}).
Figure~\ref{qz} illustrates the dependence of $q$ on redshift up to $z \sim 0.25$. For all the considered populations, $q$ remains roughly constant for the whole considered redshift range, with Seyfert galaxies having consistently lower values of $q$ than other types. A weak tendency of $q$ to decrease with redshift can be observed for Seyfert galaxies and LINERs, but the scatter of $q$ for these populations is too big to draw any firm conclusions.
At redshifts $z<0.15$ the distributions for LINERs, Seyferts and composite galaxies differ from that of the SFGs.
At the highest redshifts probed ($0.15<z<0.25$) the samples become indistinguishable, however this redshift range contains few objects (cf.~Table~\ref{ksz}). 
As shown in Fig.~\ref{qltir}, the average $q$ remains constant as a function of the FIR luminosity with a hint of slight upturn of this trend for the highest $L_{FIR}$ bin ($10^{12-12.5}$~$L_{\odot}$) for composite galaxies, Seyferts and LINERs.
K-S test shows, that  the distributions of the Seyferts and SFGs splitted into $L_{FIR}$ bins start to differ at $L_{FIR}>10^{10}$~$L_{\odot}$. 
In the highest probed $L_{FIR}$ bin (i.e. $L_{FIR}>10^{12}$~$L_{\odot}$) these galaxy samples become again indistinguishable, however this effect comes from the small number statistics (cf.~Table~\ref{ksltir}). 
At the same time, we observe a very clear decrease of the average $q$ with increasing radio power for all the considered populations of sources (see Fig.~\ref{qlradio}). 
In bins of $L_{1.4GHz}$ the K--S test shows that the $q$ distributions of SF and Seyfert galaxies are indistinguishable from each other only at the lowest radio luminosities ($L_{1.4GHz}<10^{22}$~$[W/Hz]$).
 Again, objects with the highest luminosities ($L_{1.4GHz} > 10^{24}$~$[W/Hz]$) become indistinguishable most probably due to their small number statistics (cf. Table~\ref{kslradio}).
 In sum, the $q$ distributions of SF and Seyfert galaxies differ for the highest radio ($L_{1.4GHz}\mathbf{>}10^{22}$~$[W/Hz]$) and infrared luminosities ($L_{FIR}>10^{10}$~$[W/Hz]$).
 The K--S test also shows that the composite galaxy distribution in the $10^{21}<L_{1.4GHz}/[W/Hz] <10^{23}$ range differs from the SFG sample.
 
Based on our findings, we conclude
that similar processes are behind the $q$ value decrease with the increase of the radio power 
for SFGs and galaxies containing AGNs. Alternatively, this could also mean that the line ratio diagnostics miss-classify obscured AGNs as SFG and that low $q$ values reveal the presence of obscured AGNs.


\begin{table*}
\caption{Properties of the final sample. Mean  $q$ ($<q>$) and $\sigma$ are derived from a Gaussian distribution fitting. 
}
\label{sum}
\begin{center}
\begin{tabular}{c|cccc}

\hline\hline
&SFG&Comp. & Seyferts&LINERs\\\hline
$N_{obj}$&2813&2666&392&237\\
$<q>$&$2.136\pm0.178$&$2.146\pm0.176$&$2.006\pm0.264$&$2.146\pm0.176$\\
\multicolumn{5}{c}{ }

\end{tabular}
\end{center}
\end{table*}

\begin{table*}
\caption{P-values derived from a K-S test for composite, LINERs and Seyfert galaxies under the hypothesis that they are drawn from the same population as SFGs, in redshift bins. $N$ is the total number of sources contained in a given redshift bin.}
\label{ksz}
\begin{center}
\begin{tabular}{c||c||c|c||c|c||c|c}

\hline\hline

$z$ &$N_{SFG}$&\multicolumn{2}{|c|}{Seyferts}&\multicolumn{2}{|c|}{LINERs}&\multicolumn{2}{|c}{Composites}\\\hline
&&$N$&p-values&$N$&p-values&$N$&p-values\\\hline
0.025 &2055&198 &1.34e-11& 184&0.069&1519&0.008\\
0.075 &654 & 128&4.701e-5&35 & 0.016&826 &3.199e-11\\
0.125&85& 44&0.0053& 12&0.158&233 &0.0085\\
0.175&18 & 15 &0.2809&4 &0.550&70 &0.387\\
0.225&1& 7 &0.2251&2&0.434& 18&0.461\\

\end{tabular}
\end{center}
\end{table*}
\begin{table*}
\caption{P-values derived from a K-S test for composite, LINERs and Seyfert galaxies under the hypothesis that they are drawn from the same population as SFGs, in total infrared luminosity bins. $N$ is the total number of sources contained in a given $L_{FIR}$ bin. Here we consider only sources with $L_{FIR}<10^{12.5} L_{\odot}$.}
\label{ksltir}
\begin{center}
\begin{tabular}{c||c||c|c||c|c||c|c}

\hline\hline

$\log(L_{FIR})$  &$N_{SF}$&\multicolumn{2}{|c|}{Seyferts}&\multicolumn{2}{|c|}{LINERs}&\multicolumn{2}{|c}{Composites}\\\hline
&&$N$&p-values&$N$&p-values&$N$&p-values\\\hline

\hline\hline

9.25&84&10&0.1355&5&0.332&22&0.559\\
9.75&202&17&0.531&19&0.434&60&0.013\\
10.25&646&59&3.92e-6&76&0.820&422&0.012\\
10.75&824&116&8.564e-8&68&0.494&867&0.057\\
11.25&438&102&2.548e-10&42&0.164&730&0.129\\
11.75&107&438&1.008e-6&15&0.066&353&0.995\\
12.25&5&13&0.905&8&0.644&84&0.516\\

\end{tabular}
\end{center}
\end{table*}

\begin{table*}
\caption{P-values derived from a K-S test for composite, LINERs and Seyfert galaxies under the hypothesis that they are drawn from the same population as SFGs, in radio luminosity bins. $N$ is the total number of sources contained in a given $L_{1.4GHz}$ bin. Here we consider only sources with $L_{1.4GHz}<10^{25}$~$[W/Hz]$.}
\label{kslradio}
\begin{center}
\begin{tabular}{c||c||c|c||c|c||c|c}

\hline\hline

$\log(L_{1.4 GHz})$ &$N_{SF}$&\multicolumn{2}{|c|}{Seyferts}&\multicolumn{2}{|c|}{LINERs}&\multicolumn{2}{|c}{Composites}\\\hline
&&$N$&p-values&$N$&p-values&$N$&p-values\\\hline
19.5 &28&1&0.483&2&0.027&2&0.303\\
20.5&145&11&0.581&9&0.961&44&0.045\\
21.5&923&67&0.135&95&0.706&615&4.865e-08\\
22.5&1153&188&3.43e-0.8&107&0.009&1485&5.959e-6\\
23.5&109&104&0.0005&20&0.806&385&0.010\\
24.5&4&19&0.3993&4&0.771&13&0.769\\
\multicolumn{3}{c}{ }

\end{tabular}
\end{center}
\end{table*}

\subsection{Comparison with literature}

In this section we compare our results with the previously published research focused on similar redshift ranges.
Previous such studies of the FIRC were largely based on IRAS observations, with $L_{FIR}$ calculated as the monochromatic luminosity at 60~$\mu$m.
\cite{yun01} measured the FIRC at $z<0.15$ and found $\langle q \rangle =2.34$. 
The discrepancy between the results obtained in our work and in the study of \cite{yun01} can be attributed to the difference in the total infrared luminosity estimator definitions.The $L_{FIR}$ estimator used in the work of \cite{yun01} corresponds to the integrated luminosities spanning from 42.5 to 122.5 $\mu$m, while the $L_{FIR}$ used in our work corresponds to the 80 --  170 $\mu$m wavelength range, as shown in \cite{solarz16}.
To quantify this difference we matched the AKARI-FIS $\times$ SDSS DR8 $\times$ NVSS catalogue with the IRAS faint source catalogue (FSC; \citealt{iras}) and derived the infrared luminosities using IRAS passbands following the \cite{helou88} prescription:
$L_{IRAS}=3.29\times 10^{-22}\times (2.58 L_{\nu}(60 \mu m)+L_{\nu} (100 \mu m))$~$L_{\odot}$, where $L_{\nu}$ is the luminosity per unit frequency
at a frequency $\nu=c/\lambda$, ($c$: the speed of light) and is expressed in $[erg/s/Hz]$.
For 3,593 matched sources (within a 15 arcsec matching radius) we find that on average $log(L_{IRAS}/L_{AKARI})=0.19$.
Therefore, $q_{IRAS}=log(L_{IRAS}/L_{AKARI}) + q_{AKARI}$ yields $q_{IRAS}= 2.32 $, 
which agrees with the IRAS based $q$ derived by \cite{yun01}. 
 %
 
In the work of \cite{moric10} the FIRC was explored at $z<0.2$ based on a multi-wavelength catalogue combining NVSS, SDSS and IRAS, with the $q$ parameter measured for the same four galaxy populations as in our study. 
In that work an indication of a discrepancy between the value of $q$ for Seyferts and for other galaxy classes was first discovered. \cite{moric10} derived
$\langle q \rangle = 2.14$ for Seyferts and $\langle q \rangle \approx 2.3$ for SFG, composite and LINER galaxies. 
Within uncertainty ranges, these measurements are in good agreement with our results. 


\section{Summary}

In this work we investigated the correlation of radio and far infrared emission from star-forming galaxies, composite galaxies, Seyferts and LINERs.
To this aim we used the FIR data obtained by the AKARI-FIS satellite covering the whole sky, cross-correlated with radio fluxes given by the NVSS. We identified AGNs based on the BPT diagram diagnostics making use of the SDSS DR8 data release as the spectroscopic resource.

We found that while the composite galaxy population is statistically indistinguishable from that of SFGs based on the FIRC only, Seyfert galaxies have a different distribution of the $q$ parameter quantifying this correlation.
The difference between Seyferts and SFGs is statistically significant for sources with radio and infrared luminosities in ranges $10^{12}>L_{FIR}/[L_{\odot}]>10^{10} $ and $10^{24}>L_{1.4GHz}/[W/Hz]>10^{22}$.
At the same time, the low-$q$ region ($q<1.5$) is populated mainly by Seyfert galaxies exhibiting radio excess. Furthermore, Seyfert galaxies display  bigger scatter in the FIRC than LINERs. These results are in agreement with the work of \citet{moric10}, who  first reported that the SFGs and galaxies hosting AGNs can be distinguished based on their position on the FIRC but only for the most radio-luminous sources. 

\begin{ack}
This work is based on observations with AKARI, a JAXA project with the participation of ESA.
Special thanks to Romain Thomas for the Photon \citep{photon} software and  Mark Taylor for the TOPCAT \citep{topcat} software.
A.S.  was  supported  by  the  National  Science  Centre  grant  UMO-2015/16/S/ST9/00438.
A.P\k{e}piak was  supported  by  the  National  Science  Centre  grant UMO-2014/13/N/ST9/00078.
This work was supported by the Grants-in-Aid for Scientific
Research JSPS/MEXT KAKENHI No. 17H01110.

\end{ack}

\bibliographystyle{aa}
\bibliography{solarz.bib}

\end{document}